\documentstyle[11pt,paspconf]{article}

\include{psfig}

\begin{document}

\title{Carbon stars in populations of different metallicity}
\author{M.A.T. Groenewegen}
\affil{Max-Planck-Institut f\"ur Astrophysik, 
Karl-Schwarzschild-Stra{\ss}e 1, D-85740 Garching, Germany}

\begin{abstract}
Our current knowledge of carbon stars in the Local Group and beyond,
is discussed. Although many carbon stars and late M-stars have been
identified in external galaxies a coherent understanding in terms of
the chemical evolution- and star formation rate-history of a galaxy is
still largely lacking. Issues that need to be addressed are: 1) for
some of the larger galaxies only a small fraction in area has been
surveyed so far, 2) surveys have been conducted using different techniques,
and may be incomplete in bolometric magnitude, 3) only for some
galaxies is there information about the late M-star population, 4) not
all galaxies in the Local Group have been surveyed, 5) only for a
sub-set of stars are bolometric magnitudes available.

From the existing observations one can derive the following: the
formation of carbon stars is both a function of metallicity and
star-formation. In galaxies with a similar star-formation rate
history, there will be relatively more carbon stars formed in the
system with the lower metallicity. On the other hand, the scarcity of
AGB type carbon stars in some systems with the lowest metallicity
indicates that these galaxies have had a low, if any, star-formation
rate history over the last few Gyrs.
\end{abstract}


\section{Introduction}

Carbon stars are tracers of the intermediate age population in
galaxies. Either they are currently undergoing third dredge-up (the
cool and luminous N-type carbon stars), or have been enriched with
carbon-rich material in a binary system when the present-day white
dwarf was on the AGB (the carbon dwarfs, CH-stars. The R-stars may be
the result of a coalescing binary [McClure 1997]).

Since their spectral signature is very different from oxygen-rich
stars, it is relatively easy to identify carbon stars even at large
distances.  In Sect.~2 the different techniques to identify carbon
stars are discussed, and in Sect.~3 the various surveys for carbon
stars in external galaxies are summarised. The results are discussed
in Sect.~4.

\section{Methods}

Basically three methods have been used to identify carbon stars in
external galaxies: (optical) spectroscopy of red stars, a combination
of a grating and a prism ({\sc grism}), and narrow-band filters.  The
first method was used in the ``early days'' when follow-up
spectroscopy was done on red stars identified in colour-magnitude
diagrams. This led to the discovery of the first known carbon stars in
e.g. the Carina dwarf (Cannon et al. 1980), or Fornax (Aaronson \&
Mould 1980).

Both the {\sc grism} and the narrow-band filter technique allow for a
more systematic search for carbon stars over a larger area. The former
method is predominantly used by B. Westerlund, M. Azzopardi, and
co-workers.  It aims at identifying the prominent C$_2$ band heads at
4737 and 5165 \AA. An illustration of this can be found in Westerlund
(1979). The latter method was independently used by Richer et
al. (1984) and Aaronson et al. (1984, also see Cook \& Aaronson
1989). It uses standard $V$ and $I$ filters in combination with two
narrow-band filters near 7800 and 8100 \AA, which are centered on an 
CN-band in carbon stars, and an TiO band in oxygen-rich stars,
respectively. In an [78-81] versus $[V-I]$ colour-colour plot, carbon
stars and late-type oxygen-rich stars clearly separate. For an
illustration see Cook \& Aaronson (1989), or Kerschbaum et al. (1998).

\begin{table}
\caption[]{Carbon stars in external galaxies}
\small
\begin{flushleft}
\begin{tabular}{lrrrrll} \hline
Name & D & $M_{\rm V}$ & $[$Fe/H$]$ & $N_{\rm C}$ & Area & $N_{\rm M}$ \\ 
     & (kpc) & (mag)      &            &             & (kpc$^2$) &     \\ \hline
M31  & 770 & --21.2      &   0.0      & 243         & 12   & 5254 (?) \\
Galaxy&    &--20.6      &   0.0      & 81          & 1.00 &       \\
M33  & 850 & --19.0      & --0.6      & 15          & 0.20 & 60 (3+), 15 (5+) \\
LMC  & 50 & --18.1      & --0.6      & 1045        & 4.8  & 5200(2+),1300(5+),475(6+) \\
NGC 6822 & 540 & --16.4  & --0.8      & 36          & 0.18  & 16 (3+), 1 (5+) \\
NGC 205 & 770 & --16.4   & --0.85     & 7           & 0.30  & 17 (2+)\\
SMC  & 63 & --16.2      & --1.0      & 789         & 5.4  & 1250(2+),180(5+),60(6+)\\
     &    &             &            & 1707        & 12.2 &  \\
IC 1613 & 765 & --14.9   & --1.2      & 15          & 0.52  & 6 (3+), 0 (5+)\\
WLM  & 940 & --14.0      & --1.5      & 14          & 0.28  & 6 (3+), 0 (5+)\\
Sagittarius dw & 24 & --14.0 & --0.85 & 26         &      &    \\
Fornax & 131 & --13.0     & --1.4     &104          & 0.44  & 25 (2+), 4 (5+)\\
Leo I & 270 & --12.0     &  --2.0     & 16          & 0.45   & \\
And I & 810 & --11.7      & --1.45   & 0           & & \\
And II & 700 & --11.7     & --1.3    & 8           & & 1 (?)\\
Sculptor dw & 78 & --10.7 & --1.8    & 8           & 0.63  & 40 (2+), 20 (3+)\\ 
Leo II & 230 & --10.2    &  --2.0     & 6           & 0.59  & \\
And III & 760 & --10.2   & --2.0     & 0             & & \\
Sextans & 86 & --10.0   &  --2.1     & (0)           &       & \\
Phoenix & 420 & --9.9 & --2.0       & 2            & 0.44 & \\
Carina & 87 & --9.2     & --1.5      & 9           & 0.23  & \\
Ursa Minor& 69 & --8.9  & --2.5      & 1           & 0.58 & \\
Draco  & 76 & --8.6     & --2.3      & 3           & 0.71 & \\ \hline
NGC 55   & 2150 & --20.0  & --0.6      & 14          & 3.2  & 7 (5+) \\
NGC 300  & 2150 & --18.7  & --0.4      & 16          & 3.2  & 23 (5+)\\
NGC 2403 & 3180 & --20.3  &   0.0      & 4           & 1.6  & 7 (?) \\ 
\hline
\end{tabular}
\end{flushleft}
\end{table}

\begin{figure}[t]
\centerline{\psfig{figure=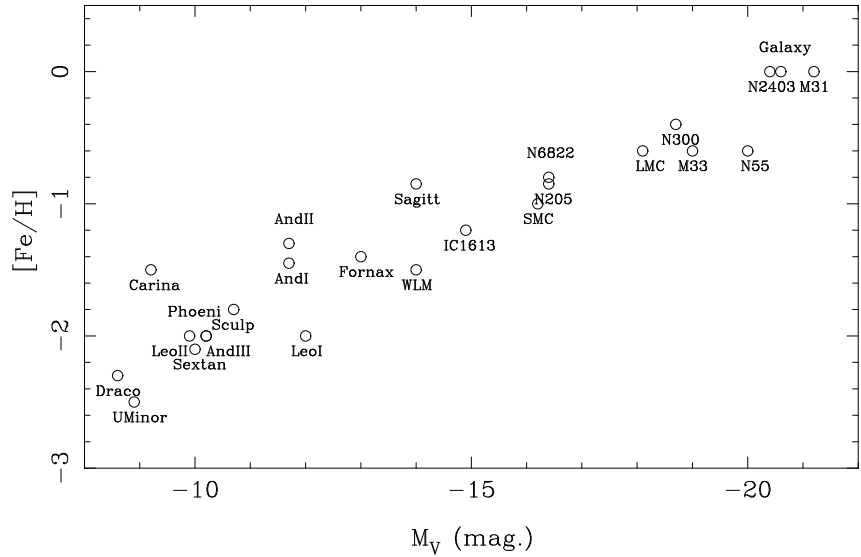,width=11.4cm}}
\caption[]{Metallicity  versus M$_{\rm V}$ in the galaxies.}
\end{figure}

\begin{figure}[t]
\centerline{\psfig{figure=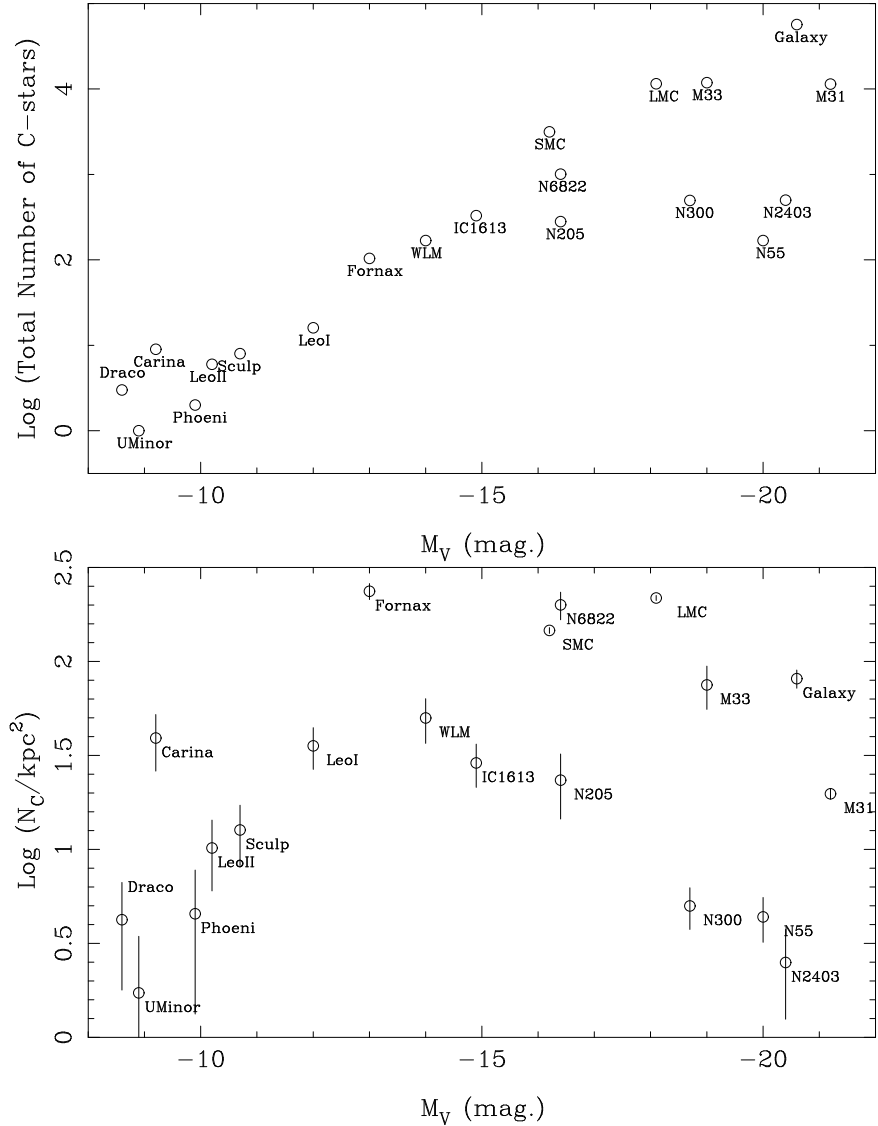,width=11.6cm}}
\caption[]{Total number of carbon stars and surface density of carbon
stars versus M$_{\rm V}$. In the latter diagram no correction for
projection effects has been made.}
\vspace{-1mm}
\end{figure}

\begin{figure}[t]
\centerline{\psfig{figure=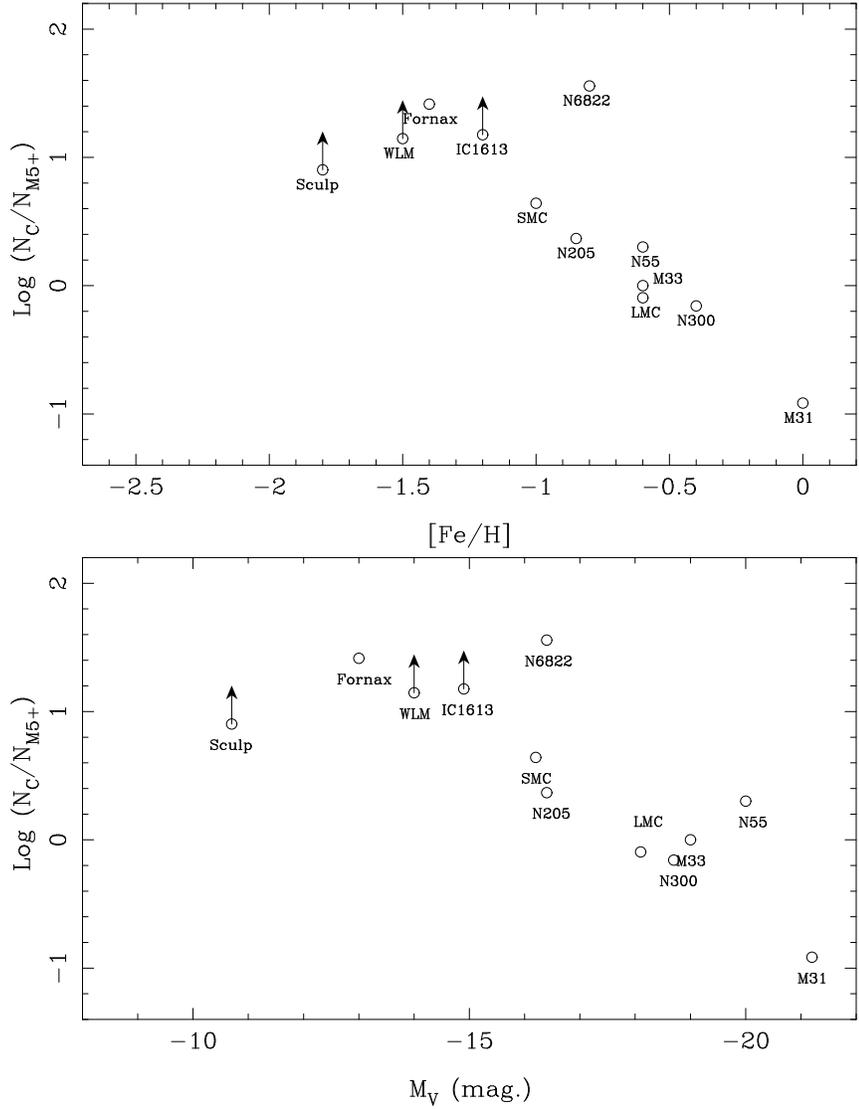,width=11.4cm}}
\caption[]{Log of the number ratio of carbon stars to late M-stars
versus metallicity and M$_{\rm V}$. }
\end{figure}

\begin{figure}[t]
\centerline{\psfig{figure=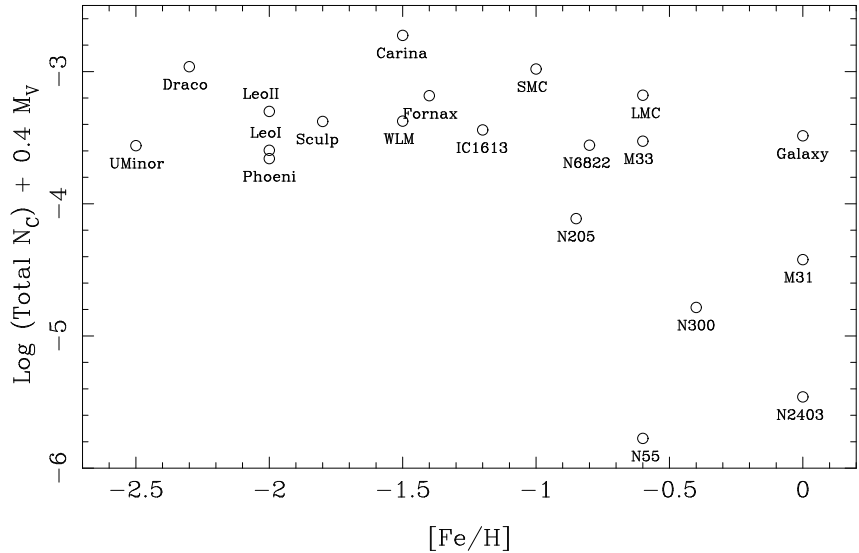,width=11.4cm}}
\caption[]{Log of the number of carbon stars over total visual
luminosity versus metallicity}
\end{figure}


\begin{figure}[t]
\centerline{\psfig{figure=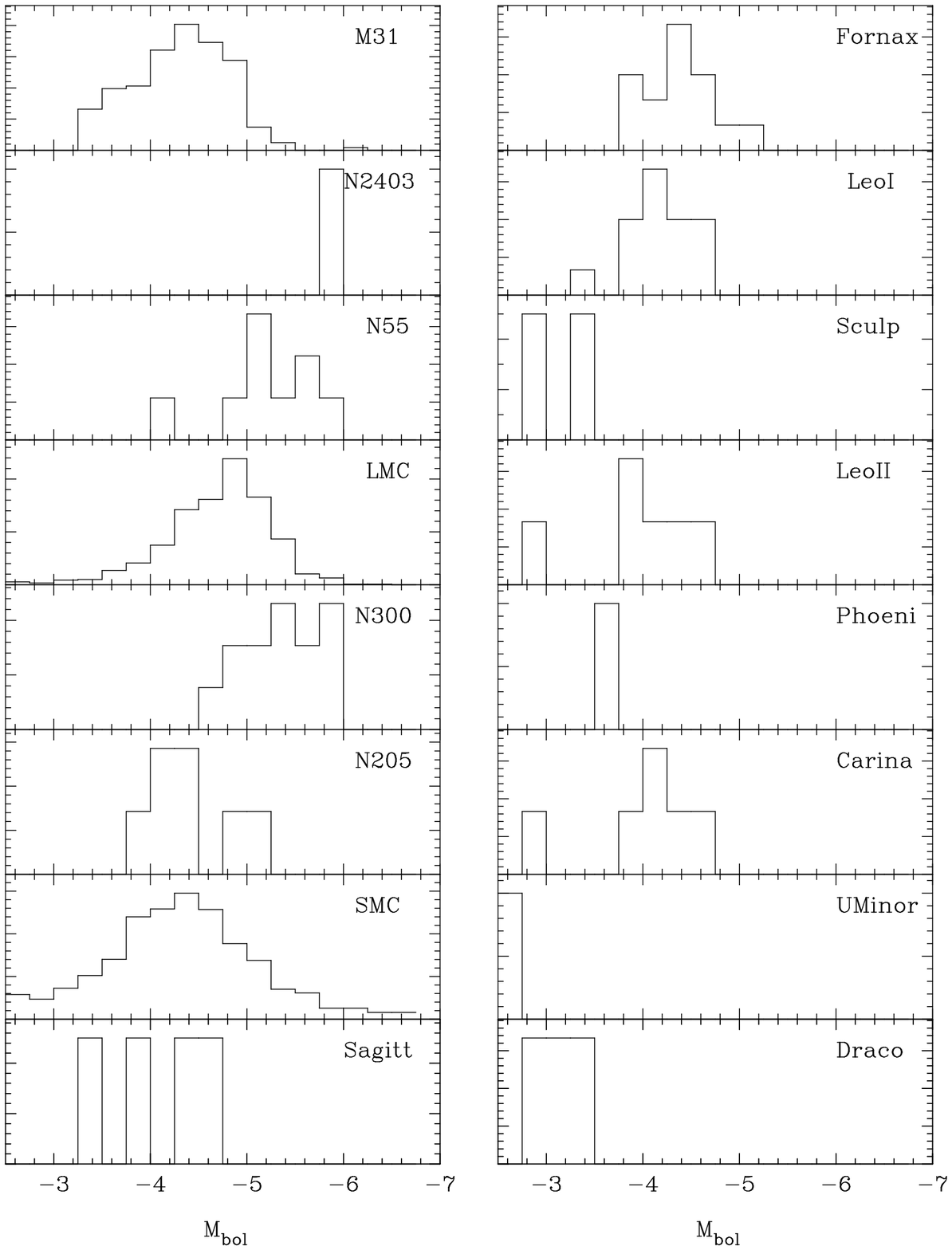,width=13.2cm}}
\caption[]{Luminosity functions, ordered by decreasing $M_{\rm
V}$. The total number of stars plotted per galaxy is listed in column
6 of Table~2. The lowest luminosity bin in the SMC and LMC are cumulative.}
\vspace{+10mm}
\end{figure}

\section{Surveys}

In this section the surveys for carbon stars in external galaxies are
described. 

\underline{The Magellanic Clouds} 

\noindent
I had the opportunity to review the subject of AGB stars in the
Magellanic Clouds  at the ``ISO's view on stellar evolution''
meeting (Groenewegen 1998) and refer to that publication for more 
details. Brief, the most complete surveys (both in area and 
magnitude) are those by Blanco \& McCarthy (1983) for both clouds, and
Rebeirot et al. (1993) for the SMC. Based on these works the total
estimated number of carbon stars in the LMC is about 11~000 and about
3~100 in the SMC.

\underline{Sagittarius dwarf galaxy}

\noindent
Ibata et al. (1995) identified 4 carbon stars which are members of the
Sagittarius dwarf. Infrared photometry for them are listed in
Whitelock et al. (1996). This data was transformed from the SAAO to
the CTIO system, and bolometric corrections from Frogel et al. (1980)
were used to derive $m_{\rm bol}$. Ng \& Schultheis (1997) suggested
another carbon star.  Since then, Whitelock (1998) reported on recent
work confirming membership for a total of 26 carbon stars, including 8
large- and 7 small-amplitude variables. This marks the discovery of
the first carbon-Mira in any of the Galactic dwarf spheroidals.

\underline{Fornax dwarf galaxy}

\noindent
Fornax has the largest number of C-stars of the galactic dwarf
spheroidals and the study of its red population dates back to Demers
\& Kunkel (1979), and Aaronson \& Mould (1980).  The total number of
known carbon stars is 104 (Azzopardi et al. 1998, and references
therein). The surveys should be area complete. Published accurate
apparent bolometric magnitudes are taken from Frogel et al. (1982) and
Aaronson \& Mould (1980) for 15 different stars, based on IR
photometry. For three stars in common the average value was taken.
Azzopardi et al. (1998) mention that they have obtained $JHK$
photometry for 94\% of these stars and find a minimum, mean and
maximum absolute bolometric magnitude of --1.2, --3.7 and --5.6,
respectively.
Frogel et al. (1982), in a smaller field, find 25 C, 6 M2+ and 0 M5+
stars. Scaling this to the total number of carbon stars I estimate
that there should be in total about 25 M2+ and possibly 4 M5+ stars in
Fornax.  Lundgren (1990) took spectra of a large number of stars in
the direction of Fornax and found 20 K-, 2 M-, 7 MS-, 7 S-, 5 SC- and
7 C-stars.

\underline{Leo {\sc i} dwarf galaxy}

\noindent
The total number of known carbon stars is 16 (Azzopardi et al. 1986,
and references therein). This survey was area complete. It should
be noted that Regulus is very close to this galaxy and therefore some
carbon stars may have escaped detection. Lee (1993)
lists the apparent bolometric magnitudes for 15 objects.

\underline{Sculptor dwarf galaxy}

\noindent
The total known number of carbon stars is 8 (Azzopardi et al. 1986,
and references therein). This survey was area complete. Apparent
bolometric magnitudes for 2 stars are listed in Frogel et al. (1982).
In a smaller area Frogel et al. (1982) found 2 C-, 10 M2+ and 5 M3+
stars. Scaling this to the total number of carbon stars results in the
number for the M-stars in Table~1.

\underline{Leo {\sc ii} dwarf galaxy}

\noindent
The total known number of carbon stars is 6 (Azzopardi et al. 1985,
and references therein). This survey was area complete. Apparent 
bolometric magnitudes  are listed in Aaronson \& Mould (1985).

\underline{Sextans dwarf galaxy}

\noindent
In the discovery paper of this galaxy by Irwin et al. (1990) it is
mentioned that there is no indication for an AGB in the
colour-magnitude diagram. On the other hand, none of the methods
mentioned in Sect.~2 was used to actively look for carbon stars.

\underline{Phoenix dwarf galaxy}

\noindent
Van de Rydt et al. (1991) provide a detailed colour-magnitude diagram
that shows a few candidate AGB stars and covers the whole of the
galaxy. Da Costa (1994) took spectra of 5 of these candidates and
confirmed two as being carbon stars. I have calculated $m_{\rm bol}$
using the photometry in van de Rydt et al., their adopted reddening
and bolometric corrections (BCs) following Bessell \& Wood (1984,
hereafter BW84).

\underline{Carina dwarf galaxy}

\noindent
The total number of known carbon stars is 9 (Azzopardi et al. 1986,
and references therein). This survey was area complete. Apparent
bolometric magnitudes are available for 6 stars from Mould et
al. (1982) based on IR photometry.

\

\newpage

\

\underline{Ursa Minor dwarf galaxy}

\noindent
The total  number of known carbon stars is 1 (Azzopardi et al. 1986,
and references therein). This survey was area complete. The apparent 
bolometric magnitude is listed in Aaronson \& Mould (1985).

\underline{Draco dwarf galaxy}

\noindent
The total known number of carbon stars is 3 (Azzopardi et al. 1986,
and references therein). This survey was area complete. Apparent 
bolometric magnitudes for the 3 stars are listed in Aaronson et al. (1982).

\underline{M31}

\noindent
This galaxy was surveyed for carbon stars by Richer \& Crabtree
(1985), Cook et al. (1986), Richer et al. (1990) and most extensively
by Brewer et al. (1995). The latter find a total of 243 C- and 5254
late-M stars (using their colour and magnitude criterion). The $V,I$
data on the individual stars is available on CD-ROM, and was used
together with their adopted reddening (depending on the field) and
BW84 to obtain $m_{\rm bol}$. The Brewer et al. survey covers about
2.1\% of the galaxy (this, and similar numbers below is derived from
the ratio between survey area and area of the galaxy, as given in the
{\sc ned} database). In a follow-up paper, Brewer et al. (1996) took
spectra of some of the stars previously identified and discovered an
S-star, and C-stars enhanced in $^{13}$C (J-type) and in Lithium.

\underline{M33}

\noindent
The total number of known carbon stars is 15 (Cook et al. 1986), who
also reported the detection of 60 late M-stars. The area surveyed
covered about 0.13\% of the galaxy. They present a LF in the $I$ band, but
no data on individual objects, so no bolometric LF could be
constructed.

\underline{NGC 205}

\noindent
This galaxy is a companion of M31. Richer et al. (1984) find 7 C and
21 late M-stars (4 of which are estimated to be foreground) in a field
that covers about 2.5\% of the galaxy. Apparent bolometric magnitudes
were derived by me from the listed $V,I$ photometry, their adopted
reddening, and BCs from BW84.

\underline{And {\sc i,ii,iii}}

\noindent
And {\sc i,ii,iii} are dwarf spheroidal companions to M31. Searches
for carbon stars are reported by Aaronson et al. (1985) for And {\sc
ii}, and by Cook, Olszewski \& Suntzeff (unpublished, quoted in
Armandroff 1994) for all three systems. Aaronson et al. (1985b) found 1
certain, 1 possible C-star and an M-giant from spectroscopy of red
stars. Cook et al. used the narrow-band filter technique to report on
no C-stars in And {\sc i} and {\sc iii}, and 8 candidate C-stars and 3
candidate S-stars in And {\sc ii}, including the one by Aaronson et
al. (1985). The area that was covered in the observations is not
quoted by Armandroff.

\underline{NGC 6822}

\noindent
Cook et al. (1986) report the discovery of 36 carbon-stars and a
similar number of late M-stars, in an area which covers about 3.5\% of
the galaxy. They present a luminosity function (LF) in the $I$ band, but no
data on the individual objects so no bolometric LF could be
derived. This galaxy also contains an S-star (Aaronson et al. 1985a)

\underline{IC 1613}

\noindent
The total number of known carbon stars is 15 (Cook et al. 1986), who
also reported the detection of 6 late M-stars. The area surveyed
covered about 4.5\% of the galaxy. They present a LF in the $I$ band, but
no data on individual objects, so no bolometric LF could be
constructed.

\underline{WLM}

\noindent
The total number of known carbon stars is 14 (Cook et al. 1986), who
also reported the detection of 6 late M-stars. The area surveyed
covered about 8.3\% of the galaxy. They present a LF in the $I$ band, but
no data on individual objects, so no bolometric LF could be
constructed.

\underline{NGC 55}

\noindent
This galaxy is in the Sculptor group. Pritchet et al. (1987) surveyed
a field covering about 8.3\% of the galaxy and found 14 C- and 7
late-M stars. The bolometric magnitude was derived by me from the
listed $V,I$ photometry, their adopted reddening and the BCs from
BW84, except for five stars with a lower limit on $V-I$.

\underline{NGC 300}

\noindent
This galaxy is in the Sculptor group. Richer et al. (1985) surveyed a
field covering about 4\% of the galaxy and found 16 C- and 23 late-M
stars. The bolometric magnitude was derived by me from the listed
$V,I$ photometry, their adopted reddening and the BCs from BW84,
except for three stars with a lower limit on $V-I$.

\underline{NGC 2403}

\noindent
This galaxy is in the M81 group, and the most distant galaxy where
carbon stars have been identified so far. Hudon et al. (1989)
tentatively identified 4 C- and 7 M-stars, and give the apparent
bolometric magnitude based on $R,I$ photometry. Their survey covered
about 0.8\% of the galaxy.

\vspace{5mm}

Table~1 summarizes the number of known carbon stars in external
galaxies. The last three entries are galaxies outside the Local
Group. Listed are the adopted distance, absolute visual magnitude,
metallicity, number of known carbon stars, the area over which they
were found, and the number of late-type M-stars, when known (the
symbol `3+' meaning stars of spectral type M3 and later, etc.). The
entry for the number of carbon stars in our galaxy is the local
surface density of TP-AGB stars by Groenewegen et al. (1992).


\begin{table}[t]
\caption[]{Carbon stars: the luminosity function}
\small
\begin{flushleft}
\begin{tabular}{lrrrrrr} \hline
Name & $M_{\rm bol}^{\rm max}$ & $M_{\rm bol}^{\rm min}$ & 
$M_{\rm bol}^{\rm mean}$ & spread & $N_{\rm all}$ & $N_{\rm M_{bol} > -3.5}$ \\
     & (mag.) & (mag.) & (mag.) & (mag.) &  &  \\ \hline
M31        & --6.14& --3.34& --4.31&  0.50&   243&   16\\
NGC 2403   & --5.91& --5.75& --5.83&  0.08&     4&    0\\
NGC 55     & --5.79& --4.18& --5.21&  0.52&     9&    0\\
LMC        & --6.30& --1.25& --4.70&  0.55&   923&   25\\
NGC 300    & --5.91& --4.66& --5.37&  0.41&    13&    0\\
NGC 205    & --5.01& --3.91& --4.38&  0.44&     7&    0\\
SMC        & --8.17& --1.62& --4.32&  0.81&  1626&  227\\
Sagittarius& --4.61& --3.29& --4.08&  0.60&     4&    1\\
Fornax     & --5.22& --3.81& --4.35&  0.37&    15&    0\\
Leo I      & --4.68& --3.38& --4.17&  0.32&    15&    1\\
Sculptor   & --3.33& --2.95& --3.14&  0.27&     2&    2\\
Leo II     & --4.51& --2.91& --3.91&  0.55&     6&    1\\
Phoenix    & --3.71& --3.55& --3.63&  0.11&     2&    0\\
Carina     & --4.55& --2.84& --4.00&  0.61&     6&    1\\
Ursa Minor & --2.74& --2.74& --2.74&  0.00&     1&    1\\
Draco      & --3.45& --2.75& --3.10&  0.35&     3&    3\\
\hline
\end{tabular}
\end{flushleft}
\end{table}

\begin{figure}[t]
\centerline{\psfig{figure=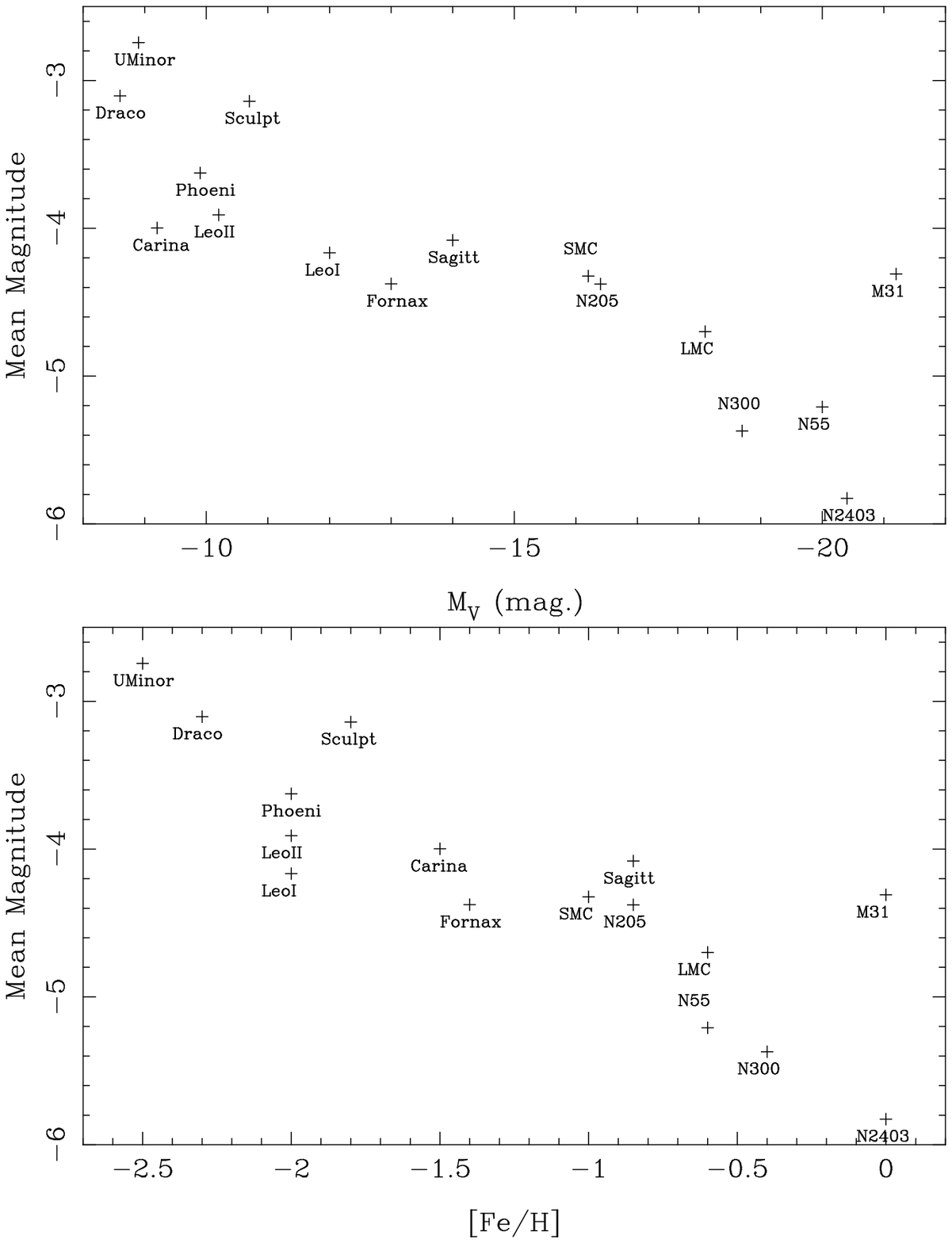,width=11.4cm}}
\caption[]{Mean bolometric magnitude of carbon stars versus M$_{\rm
V}$ and metallicity.}
\end{figure}

\section{Discussion}

Figure~1 shows the well-known correlation between metallicity and
absolute visual magnitude. The interpretation being that the more
luminous galaxies are also the more massive ones that have had more gas
mass available to transform into stars and enrich the interstellar
medium.

Figure~2 shows the number of carbon stars in a galaxy versus $M_{\rm
V}$ represented in two ways. First, the total number of C-stars in a
galaxy was estimated by multiplying the known number by the ratio of
total surface area of a galaxy to the survey area. As for some
galaxies the survey area is less than a few percent, this correction
factor can be quite large (and uncertain).  To circumvent this, the
bottom panel shows the surface density of carbon stars in the
particular survey. The drawback of this approach is that it does not
take into account the spatial variation of the density of carbon stars
within a galaxy. In neither approach did I correct for the fact that
we do not see these galaxies face-on. Some interesting things can be
noticed. There is a clear relation between the (estimated) total
number of C-stars and $M_{\rm V}$, and there seems to be a maximum
surface density of about 200 kpc$^{-2}$ averaged over a galaxy. In
both plots NGC 55, 300 and 2403 are clear outliers. These are the most
distant galaxies surveyed, and one might suppose that the surveys have
been incomplete. For NGC 55 the explanation probably lies as well in
the fact that we see this galaxy almost disk-on, and so both the total
number as well as the surface density have been
underestimated. Reddening within the galactic disk of the galaxy can
also play a role. For NGC 2403 the small number of carbon stars lies
in the fact that the survey has been incomplete. All 4 C-stars have
luminosities that are much higher than the average in galaxies for
which we know the LF is more detail. The same is true for NGC 300.  A
last note of caution is that I did not try to make a distinction
between carbon stars on the TP-AGB and lower-luminosity carbon
stars. For example, the data for our Galaxy represents TP-AGB stars
only, while on the other hand the SMC is known to contain a large
fraction of low-luminosity carbon stars (see later).

Figure~3 shows the ratio of C- to late M-stars, and it confirms the
well known trend (Cook et al. 1986, Pritchet et al. 1987). The
interpretation is that a star with a lower metallicity needs fewer
thermal pulses to turn from an oxygen-rich star into a carbon star.

Figure~4 shows the ratio of the total estimated number of carbon stars
over the visual luminosity of the galaxy (e.g. Aaronson et al. 1983). Most
of the galaxies scatter between a value of --3 and --4, with a few
outliers which are the same as noticed in Fig.~2.
 
Figure~5 shows the C-stars bolometric LF for the galaxies for which it
could be constructed (Sect.~2); for the Magellanic Clouds see
Groenewegen (1998) for details. Table~2 lists the maximum, minimum and
mean magnitude, as well as the spread, calculated from the rms
deviation from the mean.  Also listed are the number of stars that
went into the calculation and are plotted in Fig.~5, and the number of
C-stars that are fainter than $M_{\rm bol} = -3.5$. The latter is an
approximate borderline (which is in fact dependent on metallicity)
between TP-AGB and low-luminosity C-stars.  The data show that in well
populated LF, the mean $M_{\rm bol}$ is between --4 and --5. It also
shows that the mean in NGC 300 and NGC 2403 is much higher. Unless one
would invoke a large uncertainty in the distance or a burst of recent
star formation, the most natural explanation lies in the
incompleteness of the surveys in these distant galaxies.  Finally, the
data shows that in the fainter galaxies the mean magnitude increases
and that a fair number of C-stars are of the low-luminosity type. This
is more clearly seen in Fig.~6 where the mean magnitude is
plotted. This indicates a low star formation over the last few Gyrs.

In principle, the overall carbon star LF contains information about
the star-formation rate history from, say, 10 Gyr ago (the low-luminosity
C-stars in binaries) to a few-hundred Myr ago (the high luminosity
tail of the LF). Its a challenge to theoretical models to use these
constraints together with other data to derive the chemical evolution-
and star-formation rate history of these galaxies.



\begin{references}
\reference Aaronson M., Mould J., 1980, ApJ 240, 804 
\reference Aaronson M., Mould J., 1985, ApJ 290, 191 
\reference Aaronson M., Liebert J., Stocke J., 1982, ApJ 254, 507 
\reference Aaronson M., Mould J., Cook K.H., 1985a, ApJ 291, L41 
\reference Aaronson M., Olszewski E.W., Hodge P.W., 1983, ApJ 267, 271
\reference Aaronson M., Gordon G., Mould J., Olszewski E.W., Suntzeff N., 
1985b, ApJ 296, L7
\reference Aaronson M., Da Costa G.A, Hartigan P., Mould J.R., 
Norris J., Stockman H.S., 1984, ApJ 277, L9 
\reference Armandroff T.E., 1994, in: ``Dwarf galaxies'',
eds. G. Meylan and P. Prugniel, ESO, Garching, p. 211
\reference Azzopardi M., Lequeux J., Westerlund B.E., 1985, A\&A 144, 388 
\reference Azzopardi M., Lequeux J., Westerlund B.E., 1986, A\&A 161, 232 
\reference Azzopardi M., Breysacher J., Muratorio G., Westerlund B.E.,
1998, in: ``IAU symposium 192: Stellar populations in the local 
group'', in press 
\reference Bessell M.S., Wood P.R., 1984, PASP 96, 247  (BW84)
\reference Blanco V.M., McCarthy M.F., 1983, AJ 88, 1442 
\reference Brewer J., Richer H.B., Crabtree D.R., 1995, AJ 109, 2480 
\reference Brewer J., Richer H.B., Crabtree D.R., 1996, AJ 112, 491 
\reference Cannon R.D., Niss B., N$\o$rgaard-Nielsen H.U., 1980, MNRAS 196, 1P 
\reference Cook K.H., Aaronson M., 1989, AJ 97, 923 
\reference Cook K.H., Aaronson M., Norris J., 1986, ApJ 305, 634 
\reference Da Costa G.S., 1994, in: ``Dwarf galaxies'',
eds. G. Meylan and P. Prugniel, ESO, Garching, p. 221
\reference Demers S., Kunkel W.E., 1979, PASP 91, 761
\reference Frogel J.A., Persson S.E. Cohen J.G., 1980, ApJ 239, 495 
\reference Frogel J.A., Blanco V.M., McCarthy M.F., Cohen J.G.,
1982, ApJ 252, 133 
\reference Groenewegen M.A.T., 1998, in: ``ISO's view on stellar 
evolution'', eds. L.~Waters, C.~Waelkens, K.~A. van der Hucht and
P.A. Zaal, Kluwer Academic Publishers, p. 379  
\reference Groenewegen M.A.T., de Jong T., van der Bliek N.S., Slijkhuis S.,
Willems F.J., 1992, A\&A 253, 150
\reference Hudon J.D., Richer H.B., Pritchet C.J., Crabtree D.R., 
Christian C.A., Jones J., 1989, AJ 98, 1265 
\reference Ibata R.A., Gilmore G., Irwin M.J., 1995, MNRAS 277, 1354
\reference Irwin M.J., Bunclark P.S., Bridgeland M.T., McMahon R.G., 
1990, MNRAS 244, 16P
\reference Kerschbaum F., Nowotny W., Schultheis M., Hron J., 1998, in: 
``IAU symposium 192: Stellar populations in the local group'', in press 
\reference Lee M.G., Freedman W., Mateo M., Thompson I., Roth M., 
Ruiz M.-T., 1993, AJ 106, 1420 
\reference Lundgren K, 1990, A\&A 233, 21
\reference McClure R.D., 1997, PASP 109, 256
\reference Mould J., Cannon R.D., Aaronson M., Frogel J., 1982, ApJ 254, 500 
\reference Ng Y.K., Schultheis M., 1997, A\&AS 123, 115
\reference Pritchet C.J., Richer H.B., Schade D., Crabtree D.R., 
Yee H.K.C., 1987, ApJ 323, 79 
\reference Rebeirot E., Azzopardi M., Westerlund B.E., 1993, A\&AS 97, 603 
\reference Richer H.B., Crabtree D.R., 1985, ApJ 298, L13 
\reference Richer H.B., Crabtree D.R., Pritchet C.J., 1984, ApJ 287, 138 
\reference Richer H.B., Crabtree D.R., Pritchet C.J., 1990, ApJ 355, 448 
\reference Richer H.B., Pritchet C.J., Crabtree D.R., 1985, ApJ 298, 240 
\reference van de Rydt F., Demers S., Kunkel W.E., 1991, AJ 102, 130
\reference Westerlund B.E., 1979, ESO Messenger 19, 7 
\reference Whitelock P.A., 1998, in: ``IAU symposium 192: Stellar
populations in the local group'', in press
\reference Whitelock P.A., Irwin M., Catchpole R.M., 1996, New Astr. 1, 57

\end{references}
\end{document}